\documentclass[twocolumn,showpacs,preprintnumbers,amsmath,amssymb]{revtex4}

\usepackage{graphicx}
\usepackage{bm}
\begin{document}

\preprint{APS/123-QED}

\title{Transmission phase of a singly occupied quantum dot in the Kondo regime}

\author{M. Zaffalon}
\author{Aveek Bid}
\author{M. Heiblum}
\author{D. Mahalu}
\author{V. Umansky}
\affiliation{Braun Center for Submicron Research, Department of Condensed Matter Physics, Weizmann Institute of Science, 76100 Rehovot, Israel}

\date{\today}

\begin{abstract}
We report on the phase measurements on a quantum dot containing a single electron in the Kondo regime. Transport takes place through a single orbital state. Although the conductance is far from the unitary limit, we measure for the first time, a transmission phase as theoretically predicted of $\pi/2$. As the dot's coupling to the leads is decreased, with the dot entering the Coulomb blockade regime, the phase reaches a value of $\pi$. Temperature shows little effect on the phase behaviour in the range 30--600~mK, even though both the two-terminal conductance and amplitude of the Aharonov-Bohm oscillations are strongly affected. These results confirm that previous phase measurements involved transport through more than a single level.
\end{abstract}


\maketitle

What is the transmission amplitude of an electron scattered off a Kondo cloud? The Kondo effect, first observed in bulk metals doped with small concentration of magnetic impurities~\cite{hewson1997}, manifests itself as an enhancement of the scattering rate below the Kondo temperature $T_K$ (the many-body energy scale of the Kondo correlated system), below which the impurity spin is totally screened by the conduction electrons in the host metal.

The Kondo effect was predicted~\cite{ng1988, glazman1988, meir1993} and experimentally observed~\cite{dgg1998, cronenwett1998, dgg1998_2} in a quantum dot (QD), which acts as a single magnetic impurity with tunable coupling to the screening electrons in the leads. In fact, a QD~\cite{kouwenhoven1997, hanson2007}, a small confined region connected by two tunnel barriers to electron reservoirs, is characterised by an on-site charging energy $U$ owing to its small capacitance; by level quantisation $\epsilon$ because of lateral confinement; and by homogeneous level broadening due to finite coupling, $\Gamma$. In addition, the QD energy levels can be tuned, allowing to change the QD occupancy $\langle N \rangle$.

The Kondo effect in a QD is easily probed by conductance measurements. Conductance takes place approximately at the charge degeneracy points, involving, say, electrons 0 and 1, when the energy $\epsilon_0$ to add the first electron to the empty dot is $\epsilon_0 \approx 0$ (or when $\epsilon_0 + U \approx 0$, to add a second electron with opposite spin). Away from these points only cotunneling, processes involving two or more simultaneous tunnelling events, occur and the conductance is expected to be suppressed. However, when $\langle N\rangle=1$, the Kondo effect allows for a substantial current flow if $T \lesssim T_K$, reaching a maximum conductance of $G_{max} = 2e^2/h$, the unitary limit, at $T=0$. Strictly speaking, the Kondo regime is limited to the the parameter region $-U + \Gamma \lesssim \epsilon_0 \lesssim - \Gamma$~\cite{dgg1998_2, gerland2000}. If $\Gamma / \Delta\epsilon \lesssim 0.5$, one single orbital state is involved and $T_K(\epsilon_0) = \frac{\sqrt{\Gamma U}}{2}\cdot\exp\left (\frac{\pi\epsilon_0(\epsilon_0 + U)}{\Gamma U}\right )$: with this definition $G(T_K) = G_{max}/2$, with $G_{max} = 2e^2/h$ if the barriers are symmetric~\cite{dgg1998_2}. In the following we consider single level transport.

Whereas the conductance is proportional to the transmission probability through the QD, it discards informations about the electron transmission phase. At $T = 0$ the conductance $G$ and the transmission phase $\delta$ are related by $G = G_{max}\sin^2 \delta$~\cite{nozieres1974, dgg1998_2, pustilnik2004}, predicting a monotonic phase evolution with a phase rise of $\pi /2$ from $\langle N \rangle = 0$ to $\langle N \rangle = 1$, followed by a constant phase shift in the Kondo valley ($\epsilon_0 =-U/2$), raising to $\pi$ when the QD is doubly occupied, outside the Kondo region. At finite temperature, the monotonicity disappears, but even at $T \approx 10\ T_K$, the transmission phase climbs only to $0.7\pi$ across the first peak before decreasing to $\pi/2$ in the Kondo valley~\cite{silvestrov2003}. At $T \gg T_K$, Kondo correlations are negligible and the well known result for the Coulomb regime is expected: a $\pi$ rise across each peak~\cite{hackenbroich1996, mak2005} and a phase lapse in the conductance valley, with a \emph{total} phase evolution of $\pi$. For an arbitrary temperature, the phase can be calculated only by numerical renormalisation group~\cite{gerland2000}: these results, based on the single level Anderson impurity model, predict a smooth evolution between the Kondo and the Coulomb blockade regime.

Previous measurements of the phase shift by Ji {\it et al.}~\cite{ji2000, ji2002} were obtained using an interferometer with the Kondo-dot placed in one of its arms. They found a phase evolution of about $1.5\pi$ across the two peaks, both in the non-unitary and in the unitary limit, the two phase evolutions differing only in the presence of a plateau situated at $\delta = \pi$ in the Kondo valley in the non-unitary case. A different approach was taken by Sato {\it et al.}~\cite{sato2005} who studied the resonances induced by a Kondo-dot side coupled to a quantum wire. They deduced a phase shift of $\pi / 2$ by analysing Fano resonances in the conductance.

\begin{figure}
\includegraphics[width=0.22\textwidth]{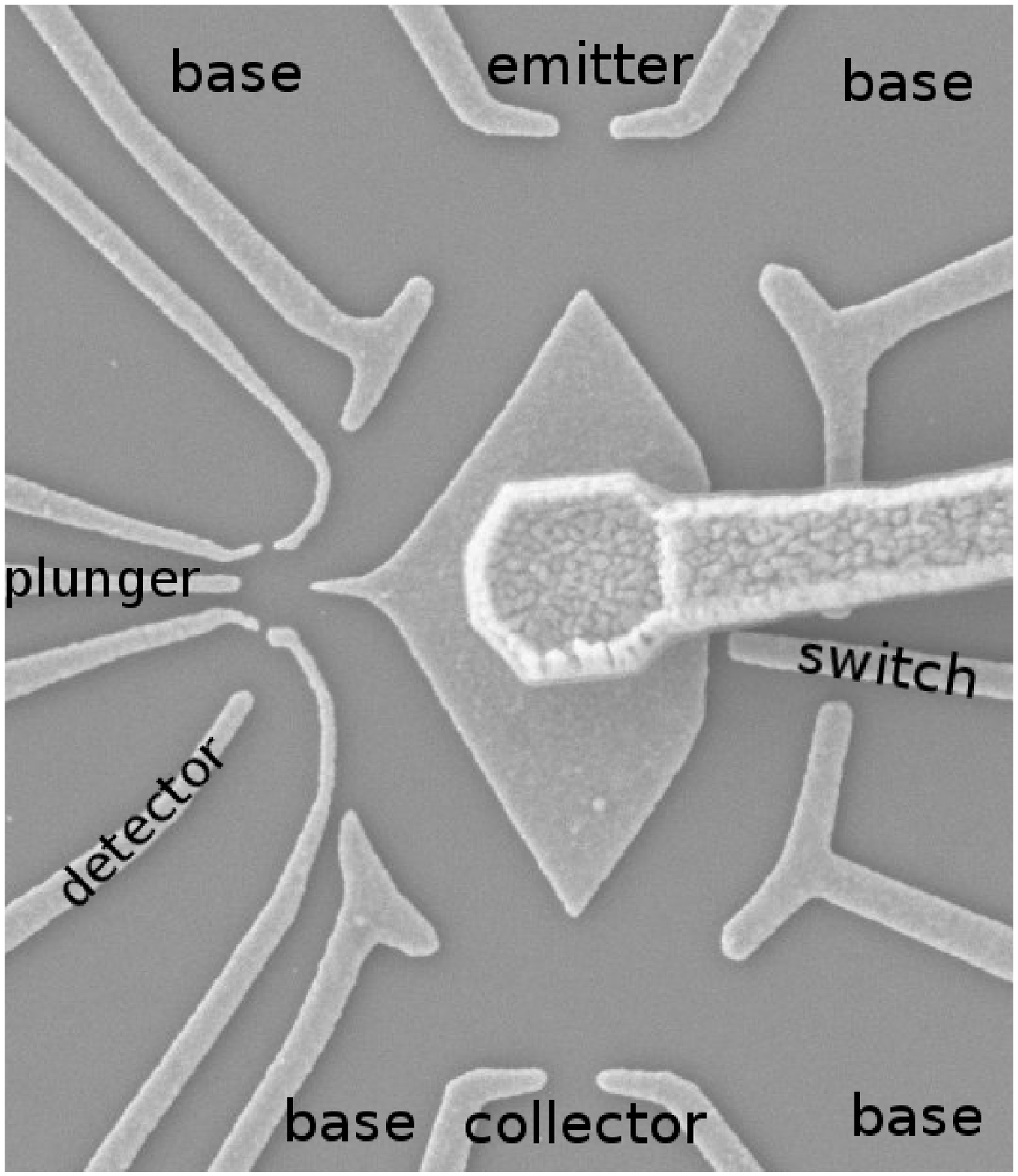}\includegraphics[width=0.28\textwidth]{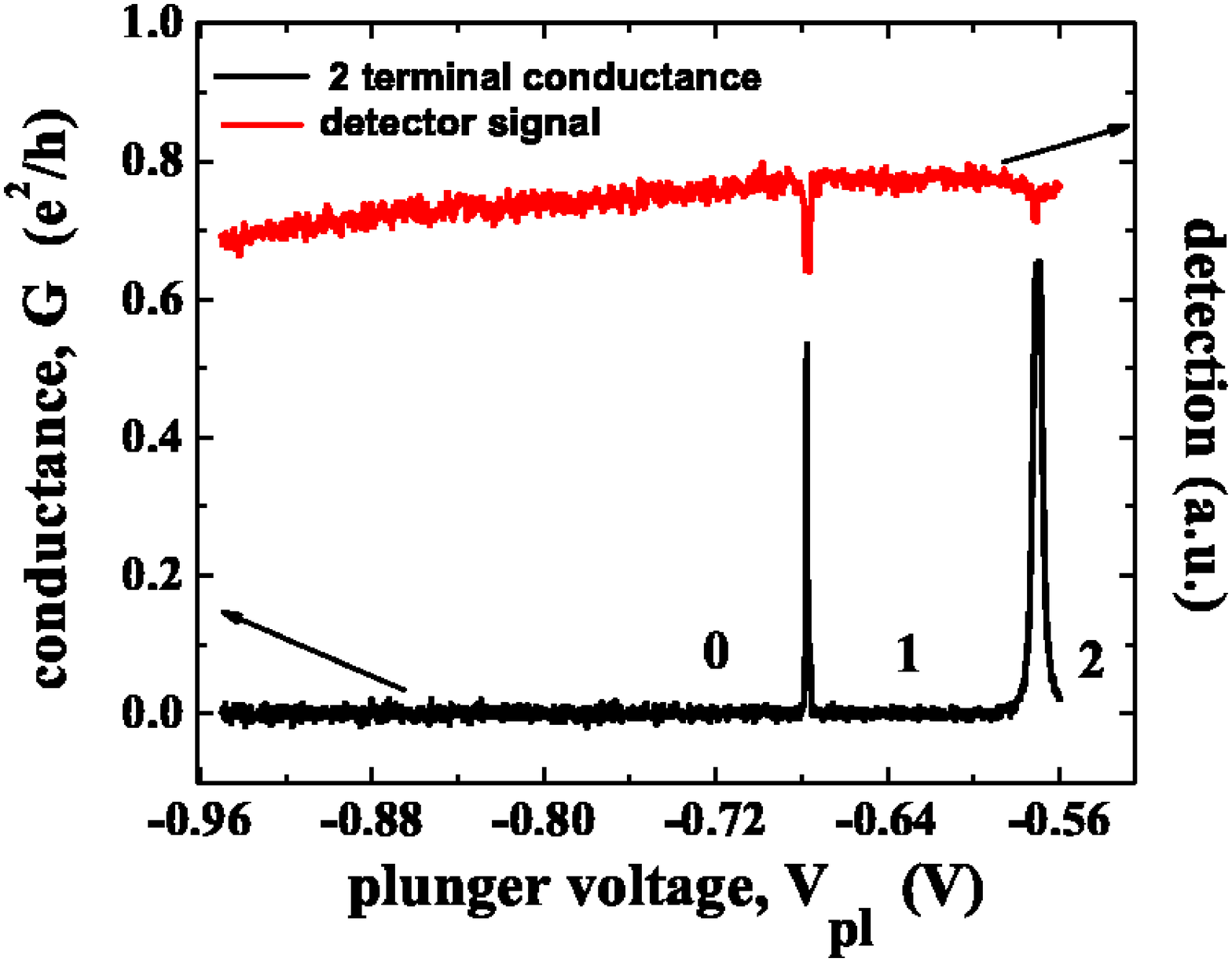}
\caption{SEM micrograph of the device and electron counting. Left: Micrograph of the interferometer with the QD embedded in one arm. Right: Two-terminal conductance of the QD with the last two electrons and the detector signal.}
\label{fig:fig1}
\end{figure}

There is by now vast theoretical evidence (for a summary, see~\cite{hackenbroich2001}) that the transmission phase depends on the specific properties of the QD's levels that participate in the transport. In this Letter we report on measurements on a QD in which transport takes place through a single level. To ensure this condition, we fabricated a QD, similar to the type described by Ciorga {\it et al.}~\cite{ciorga2000} in which transport is still substantial when $\langle N\rangle=1$ electron~\cite{vidan2006}. This gives large level spacings, as the \emph{average} level spacing decreases with $1/\sqrt{\langle N\rangle}$. Even so, for a typical QD, $U=3.0$~meV and $\Delta \epsilon = 0.8$~meV, a small Kondo temperature of $15-20$~mK is predicted at $\epsilon_0 = -U/2$.

The determination of the phase evolution is based on the interference between two paths - the two arms of an interferometer - one of which contains the QD~\cite{yacoby1995} and the other being the reference arm, with transmission amplitudes $t_{QD}^{coh}(\epsilon_0) = |t_{QD}^{coh}(\epsilon_0)|e^{i \varphi_{QD}(\epsilon_0)}$ and $t_{ref}$, respectively. In an \emph{open} interferometer~\cite{schuster1997, ji2000, mak2005}, four grounded bases collect the backscattered electrons and only two direct paths from emitter to collector are possible. A weak magnetic field $B$ threading the island adds an Aharonov-Bohm (AB) phase $\varphi_{AB} = 2\pi\phi / \phi_0$ to the electron, where $\phi$ is the magnetic flux enclosed by the electron path and $\phi_0=h/e$ is the flux quantum. The coherent current at the collector is proportional to $|t_{ref} + t_{QD}^{coh}e^{-i\varphi_{AB}}|^2 =$ (constant term) $+ 2|t_{ref}||t_{QD}^{coh}|\cos(\varphi_{AB} + \varphi_{QD} + \phi')$: the former part is weakly $B$-dependent, owing to the Lorentz force and the latter, $T^{flux}(\epsilon_0)\cos\left [\varphi_{AB} + \varphi_{QD}(\epsilon_0) + \phi'\right ]$ is periodic in the flux quantum; $\phi'$ is a constant interferometer-dependent phase.

Referring to Fig.~\ref{fig:fig1}, the device is fabricated on a two dimensional electron gas embedded in a AlGaAs/GaAs heterostructure, some 60~nm beneath the surface, with carrier density of $3.3\times 10^{15}$~m$^2$ and mobility of $1.2\times 10^{2}$~V/m$^2$s at 4.2~K. Two subsequent steps of electron beam lithography are required to pattern the gates and the bridge. The four reflectors can be individually biased in order to focus the electrons from emitter to collector and increase the signal. The reference arm, which can be blocked by the switch gate, carries approximately 10 conducting modes and the arm with the dot about 5. Measurements were performed on one device in a dilution refrigerator with electron temperature of 30~mK. Conductance measurements of the QD were taken with $v_{sd} = 5~\mu$V excitation voltage below 300~mK and $v_{sd} = 10-20~\mu$V above 300~mK at 250~Hz and the current was measured with an Ithaco 1211 current preamplifier. A second device, measured in a dilution refrigerator with electron temperature of 150~mK behaved in a quantitative similar manner.

\begin{figure}
\includegraphics[width=0.45\textwidth, height=0.32\textwidth]{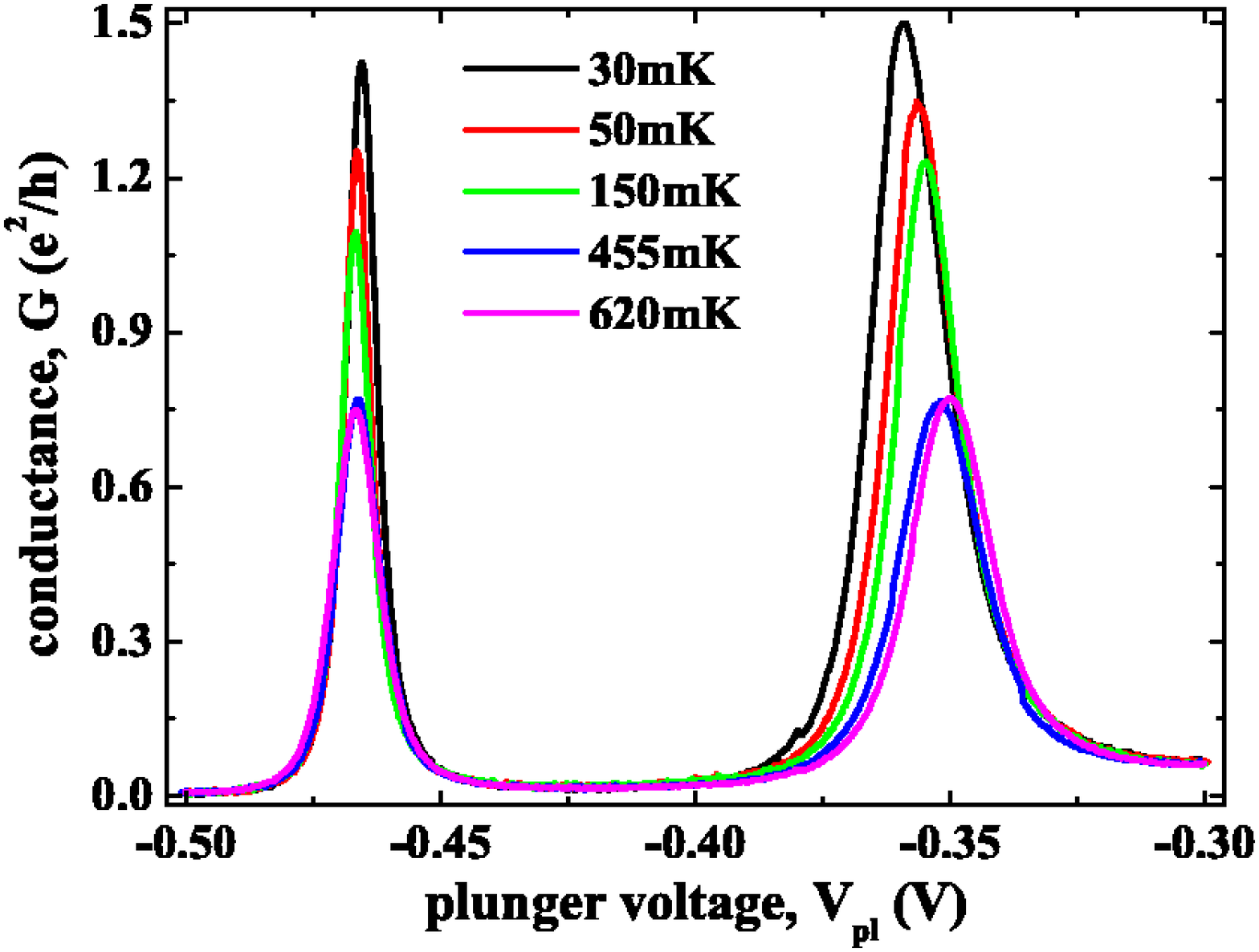}
\includegraphics[width=0.45\textwidth, height=0.32\textwidth]{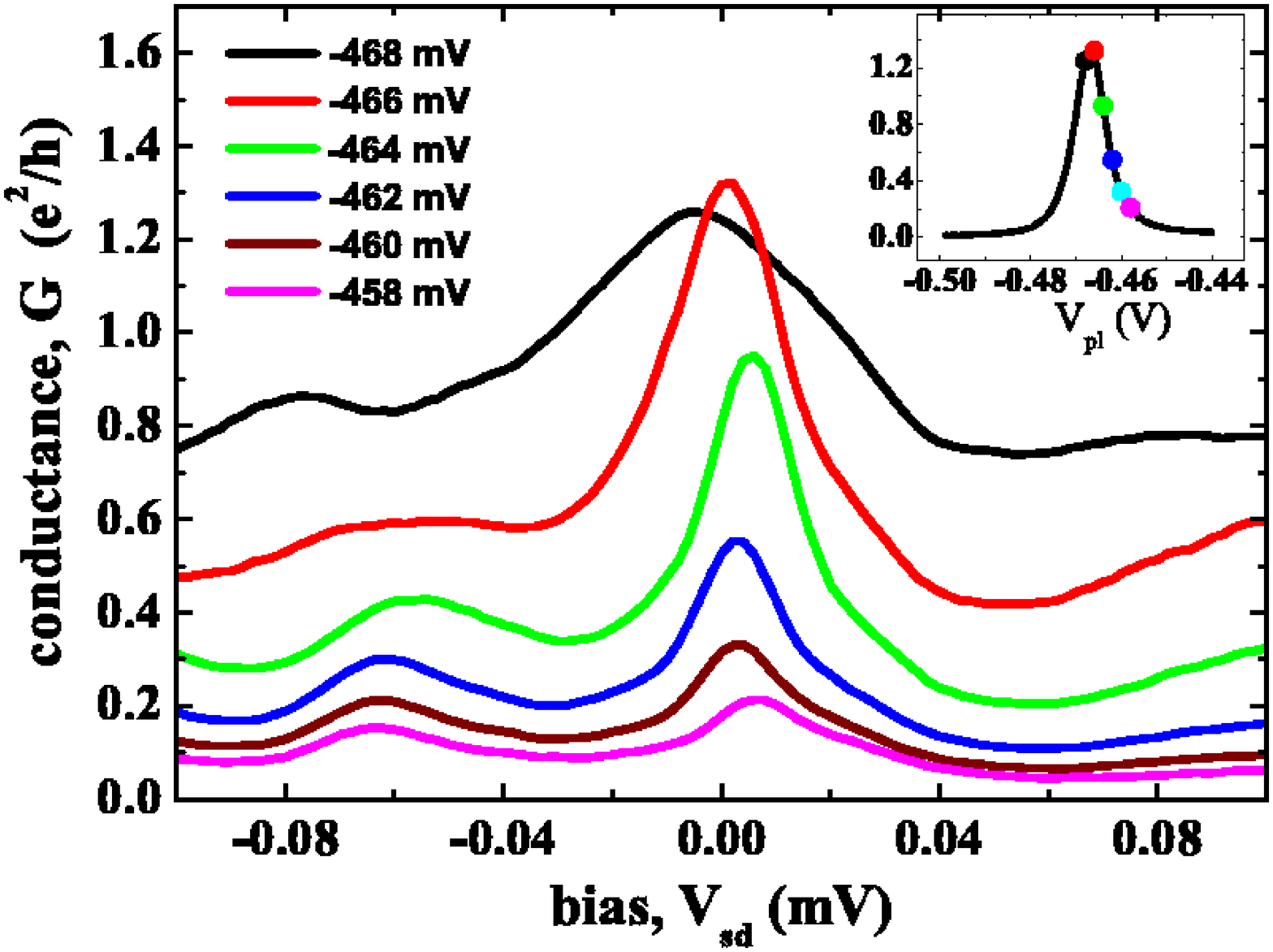}
\caption{Two-terminal conductance: temperature and bias dependence. Top: Temperature dependence of the first two peaks ($\Gamma = 180\pm 25~\mu$eV). Bottom: Two-terminal finite source drain bias scans taken in correspondence to the first peak at plunger biases indicated by the dots on the trace in the inset.}
\label{fig:fig2}
\end{figure}

A quantum point contact (QPC) is situated in close proximity to the QD. It detects the average QD occupation, as the conductance through the QPC is affected by the electrostatic potential of the QD. In order to enhance the sensitivity of the detection, we employ the measuring scheme previously used by Sprinzak {\it et al.}~\cite{sprinzak2002}. Fig.~\ref{fig:fig1}(b) shows the dot's two-terminal conductance, together with the detector signal, revealing the QD \emph{absolute} occupancy $\langle N\rangle=0,1,2$: a dip in the detector signal appears whenever the QD average occupation changes by one electron

Two-terminal conductance measurements of the QD are taken between the bases in order to avoid the emitter and collector series resistance. From finite bias scans, we evaluate the charging energy to be $U\approx 3.0\pm 0.2$~meV and the first excited state for the first electron to be $\Delta \epsilon_0 = 0.80\pm 0.05\ \mu$eV, whereas for the second peak, the level spacing decreases to $\Delta \epsilon_1 = 0.30\pm 0.05\ \mu$eV. We then tune the coupling of the QD to the leads so as to maximise the conductance of the first two peaks, with the constrain that the conductance in the $\langle N\rangle=2$ valley is lower than about $0.05-0.1e^2/h$: failing to do so results in the phase evolution being featureless, probably due to multilevel transport. On the offside, this results in a lower $T_K$ in the $\langle N\rangle=1$ valley.

The presence of Kondo correlations is verified by the following characteristics: the peaks' conductance being larger than $G_0 = e^2/h$ and the suppression of the peak conductance by the application of a finite bias~\cite{dgg1998_2}. Fig.~\ref{fig:fig2}(a) shows a series of two terminal conductance traces of the first two peaks, as a function of the temperature. Differential conductance traces for some plunger voltages are reported in Fig.~\ref{fig:fig2}(b) at base temperature: the resonances at finite bias are probably induced by the reflectors.

\begin{figure}
\includegraphics[width=0.5\textwidth, height=0.35\textwidth]{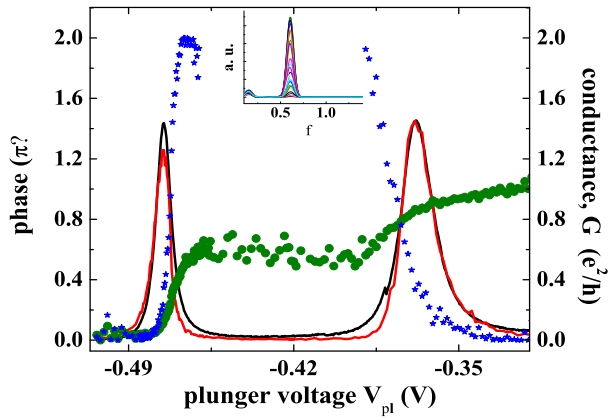}
\vspace{-0.5cm}
\caption{Phase evolution in the QD. Two-terminal conductance (black trace), AB amplitude squared, $(T^{flux})^2$ (red trace), and phase of two peaks (green dots). The turquois stars are calculated from the expression $G = 2e^2/h \sin^2\delta$. Inset: Amplitude squared of the AB oscillations, showing $f_0$ the fundamental frequency.}
\label{fig:fig3}
\end{figure}

Once the Kondo-enhanced peaks are identified, we set the emitter and collector QPCs of the interferometer to a conductance of $2-3~e^2/h$. We then open the reference arm and measure the ballistic current between emitter and collector with all bases grounded, while a weak magnetic field is swept, in the range of tens of mT. Typically, about 95\% of the injected current is lost to the bases. The current at the collector shows AB oscillations with visibility (ratio between amplitude of the oscillations and average background) of about 20\%. The period $f_0 = 0.61~\mathrm{mT}^{-1}$ corresponds to an area enclosed by the electron paths of $1.64~\mu\mathrm{m}^2$, comparable to the interferometer area of $1.7~\mu\mathrm{m}^2$.

The amplitude squared $(T^{flux})^2 \propto |t_{QD}^{coh}|^2$ of the AB oscillations at the frequency $f_0$ is plotted in Fig.~\ref{fig:fig3}, together with the phase evolution, as determined by Fourier analysis and normalised such that the maximum of $(T^{flux})^2$ coincides with the maximum of $G_{2t}$, the red trace in the Figure. This is based on the assumption that at $T \ll T_K$ all transport processes are coherent~\cite{gerland2000, nozieres1974}.

\begin{figure}
\includegraphics[width=0.5\textwidth]{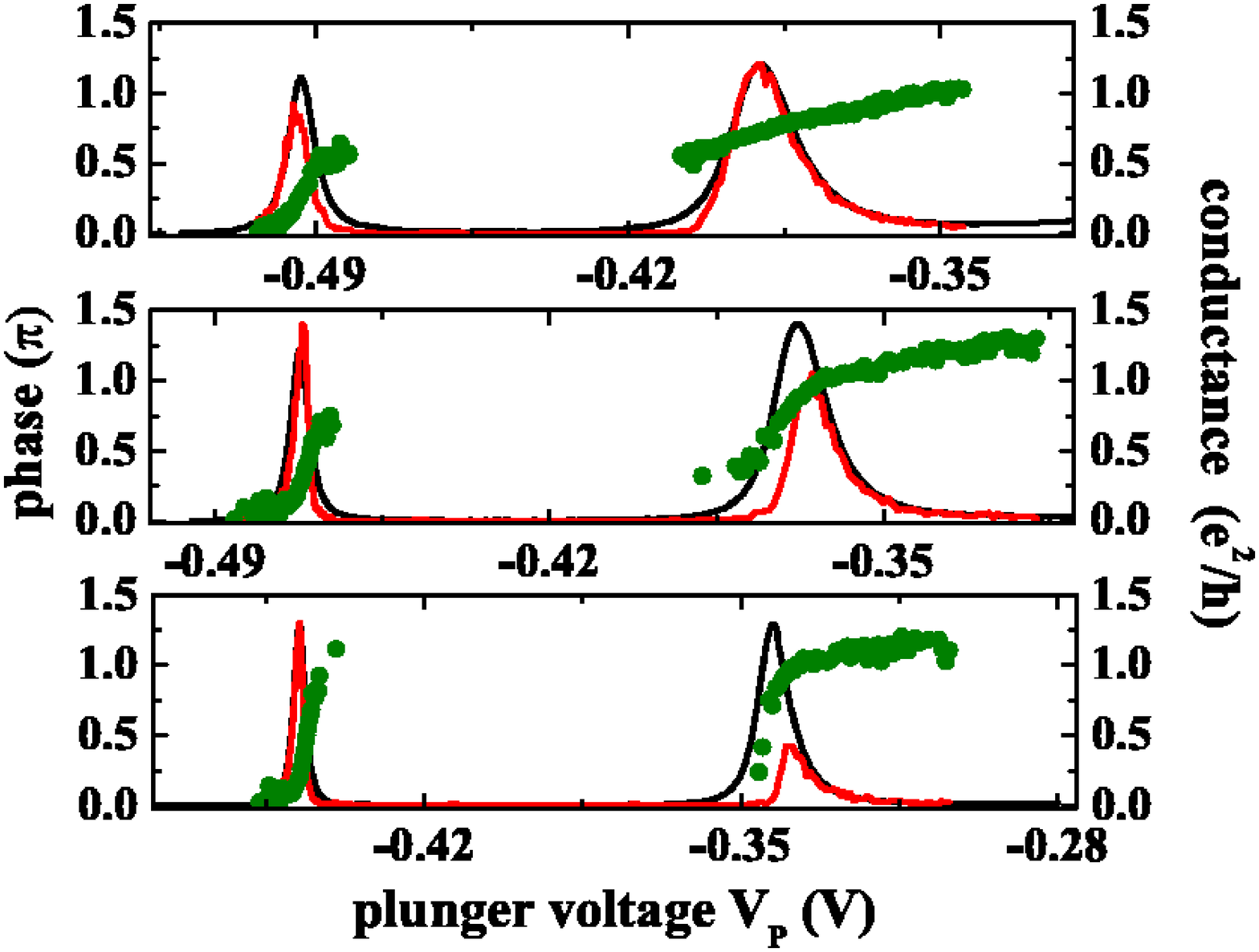}
\caption{Phase evolution for $\Gamma=180,\ 110$ and 50~$\mu$eV at $T=30$~mK. $(T^{flux})^2$ has been normalised to the $G_{2t}$, even though at smaller $\Gamma$, the above mentioned argument ceases to be valid as $T \gtrsim T_K$.}
\label{fig:fig4}
\end{figure}

Three features are evident: a) the phase evolution across the two peaks is $\pi$ and it seems to saturate at a value of $\approx 0.5\pi$ across the first peak and again $\approx 0.5\pi$ across the second peak, even though for the second peak the condition of single-level transport $\Gamma / \Delta \epsilon_2 \lesssim 0.5$ is not strictly satisfied. In the valley, the coherent current is below the noise level and the phase evolution can barely be followed. This phase behaviour is in qualitative agreement with the prediction of Gerland {\it et al.}~\cite{gerland2000} for transport through one orbital level and in disagreement with the previous measurement in a similar open interferometer of Ji {\it et al.}~\cite{ji2000}. b) $(T^{flux})^2$ closely follows $G_{2t}$ except in the $\langle N\rangle=1$ conductance valley: in fact $G_{2t}$ includes also incoherent processes that do not contribute to the AB oscillations. An example of such processes are cotunneling events accompanied by spin-flips: for instance, a spin-up electron in the QD tunnels out and is replaced by a spin down electron from the lead~\cite{konig2002, silvestrov2003}. c) The turquois stars in Fig.~\ref{fig:fig3} are calculated from the relationship $G = G_{max}\sin^2 \delta$, valid outside the Kondo regime where $T \ll \Gamma / k_B$, $\Gamma$ being there the lowest energy scale. Here we set $G_{max} = 2e^2/h$, as the couplings are approximately equal. It is evident that a different choice of $G_{max}$ would not give a better agreement.

A first conclusion can now be drawn: although the QD exhibits Coulomb blockade-like features (highly suppressed current in the valley) and only energy and temperature dependence reveal Kondo correlations, the phase evolution is drastically different to that in the Coulomb regime, proving the extreme sensitivity of the phase to correlations~\cite{silvestrov2003}.

We now proceed to decrease the QD coupling so as to decrease $T_K$. We expect to see a transition to a $\pi$ phase shift across each peak in the Coulomb blockade limit. Fig.~(\ref{fig:fig4}) shows such transition of the phase evolution to $0.9\pi$ when the first peak width is about 50~$\mu$eV. In contrast to the results reported by Avinun-Kalish {\it et al.}~\cite{mak2005}, the phase across the first two peaks is limited to $\pi$ rise, indicating transport through the same orbital~\cite{hackenbroich2001}.

\begin{figure}
\includegraphics[width=0.5\textwidth]{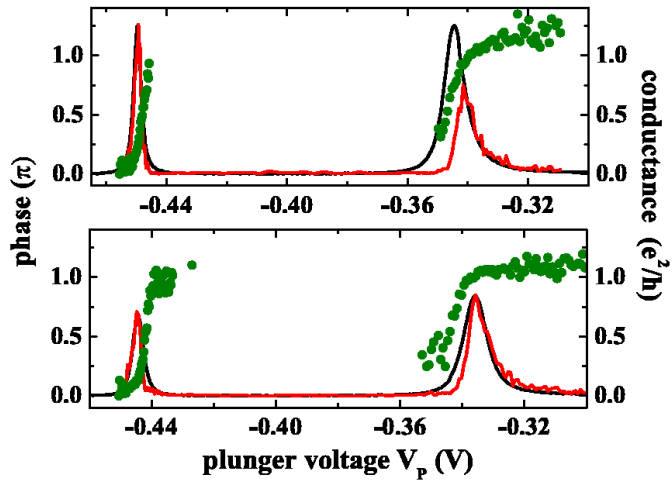}
\caption{Phase evolution as a function of the temperature for the narrowest peak, $\Gamma = 80\pm10\ \mu$eV at base temperature and 300~mK, showing a transition to $\pi$ phase evolution across each peak.}
\label{fig:fig5}
\end{figure}

The Kondo effect can also be suppressed by raising the temperature: we now proceed to measure the the temperature dependence of the phase, in the temperature range $30-600$~mK, Fig.~\ref{fig:fig5}. The measurements are taken in the following way: we tune the peaks at base temperature to have the required width, and scan the two-terminal and multi-terminal conductances at different temperatures. A little re-tuning is sometimes required between one temperature value and the following as a typical phase scan requires approximately 10~hours.

Both $G_{2t}$ and the AB oscillations are strongly affected by temperature: $(T^{flux})^2$ at the highest temperature is $\approx 30$ times smaller than that at base temperature. However as for the phase, no temperature dependence  is  observed for the wide peaks ($\Gamma = 180$ and $\Gamma = 110\ \mu$eV), up to 600~mK within the experimental error. For the narrow peak, $\Gamma = 80~\mu$eV, the phase is $0.8\pi$ at base temperature and increases to $\pi$ at $T=300$~mK. This is consistent with the robustness of the phase evolution pointed out by Silvestrov~{\it et al.}~\cite{silvestrov2003}.

In conclusion we have shown that the transmission phase through a QD evolves from a $\pi$ phase shift in the Coulomb regime to $\approx \pi/2$ in the Kondo and it persists at temperatures up to $5-10$ times $T_K$. A temperature induced change of the phase evolution could only be seen with the smallest coupling. These results provide some more insight in the previously measured phase evolution through a QD. We are in the position to identify three distinct behaviours: 1) $\Gamma \approx k_BT$ gives the Coulomb blockade result $\pi$, for the transmission phase; 2) $\Gamma \gtrsim 30k_BT$, the Kondo result of $\pi/2$; 3) $\Gamma \gg \Delta \epsilon$, i.e. multilevel transport~\cite{schuster1997, mak2005}, a $\pi$ phase rise across each peak.

We acknowledge useful discussions with Yuval Gefen, Keren Michaeli, Peter Silvestrov, Theresa Hecht, Jan von Delft, Andrei Kretinin, Yaron Bromberg and especially Yuval Oreg and David Goldhaber-Gordon and the technical help of Sandra Foletti, Yunchul Chung, Oren Zarchin and Michal Avinun-Kalish.

\bibliography{kondo_phase}

\begin{thebibliography}{25}
\expandafter\ifx\csname natexlab\endcsname\relax\def\natexlab#1{#1}\fi
\expandafter\ifx\csname bibnamefont\endcsname\relax
  \def\bibnamefont#1{#1}\fi
\expandafter\ifx\csname bibfnamefont\endcsname\relax
  \def\bibfnamefont#1{#1}\fi
\expandafter\ifx\csname citenamefont\endcsname\relax
  \def\citenamefont#1{#1}\fi
\expandafter\ifx\csname url\endcsname\relax
  \def\url#1{\texttt{#1}}\fi
\expandafter\ifx\csname urlprefix\endcsname\relax\def\urlprefix{URL }\fi
\providecommand{\bibinfo}[2]{#2}
\providecommand{\eprint}[2][]{\url{#2}}

\bibitem[{\citenamefont{Hewson}(1997)}]{hewson1997}
\bibinfo{author}{\bibfnamefont{A.~C.} \bibnamefont{Hewson}},
  \emph{\bibinfo{title}{The Kondo Problem to Heavy Fermions}}
  (\bibinfo{publisher}{Cambridge University press}, \bibinfo{year}{1997}).

\bibitem[{\citenamefont{Ng and Lee}(1988)}]{ng1988}
\bibinfo{author}{\bibfnamefont{T.~K.} \bibnamefont{Ng}} \bibnamefont{and}
  \bibinfo{author}{\bibfnamefont{P.~A.} \bibnamefont{Lee}},
  \bibinfo{journal}{Phys. Rev. Lett.} \textbf{\bibinfo{volume}{61}},
  \bibinfo{pages}{1768} (\bibinfo{year}{1988}).

\bibitem[{\citenamefont{Glazman and Raikh}(1988)}]{glazman1988}
\bibinfo{author}{\bibfnamefont{L.~I.} \bibnamefont{Glazman}} \bibnamefont{and}
  \bibinfo{author}{\bibfnamefont{M.~E.} \bibnamefont{Raikh}},
  \bibinfo{journal}{JETP Lett.} \textbf{\bibinfo{volume}{47}},
  \bibinfo{pages}{452} (\bibinfo{year}{1988}).

\bibitem[{\citenamefont{Meir and Wingreen}(1993)}]{meir1993}
\bibinfo{author}{\bibfnamefont{Y.}~\bibnamefont{Meir}} \bibnamefont{and}
  \bibinfo{author}{\bibfnamefont{N.~S.} \bibnamefont{Wingreen}},
  \bibinfo{journal}{Phys. Rev. Lett.} \textbf{\bibinfo{volume}{70}},
  \bibinfo{pages}{2601} (\bibinfo{year}{1993}).

\bibitem[{\citenamefont{Goldhaber-Gordon
  et~al.}(1988)\citenamefont{Goldhaber-Gordon, Shtrikman, Mahalu, Abush-Magder,
  Meirav, and Kastner}}]{dgg1998}
\bibinfo{author}{\bibfnamefont{D.}~\bibnamefont{Goldhaber-Gordon}},
  \bibinfo{author}{\bibfnamefont{H.}~\bibnamefont{Shtrikman}},
  \bibinfo{author}{\bibfnamefont{D.}~\bibnamefont{Mahalu}},
  \bibinfo{author}{\bibfnamefont{D.}~\bibnamefont{Abush-Magder}},
  \bibinfo{author}{\bibfnamefont{U.}~\bibnamefont{Meirav}}, \bibnamefont{and}
  \bibinfo{author}{\bibfnamefont{M.~A.} \bibnamefont{Kastner}},
  \bibinfo{journal}{Nature} \textbf{\bibinfo{volume}{391}},
  \bibinfo{pages}{156} (\bibinfo{year}{1988}).

\bibitem[{\citenamefont{Cronenwett et~al.}(1998)\citenamefont{Cronenwett,
  Oosterkamp, and Kouwenhoven}}]{cronenwett1998}
\bibinfo{author}{\bibfnamefont{S.~M.} \bibnamefont{Cronenwett}},
  \bibinfo{author}{\bibfnamefont{T.~H.} \bibnamefont{Oosterkamp}},
  \bibnamefont{and} \bibinfo{author}{\bibfnamefont{L.~P.}
  \bibnamefont{Kouwenhoven}}, \bibinfo{journal}{Science}
  \textbf{\bibinfo{volume}{281}}, \bibinfo{pages}{540} (\bibinfo{year}{1998}).

\bibitem[{\citenamefont{Goldhaber-Gordon
  et~al.}(1998)\citenamefont{Goldhaber-Gordon, G{\"o}res, Kastner, Shtrikman,
  Mahalu, and Meirav}}]{dgg1998_2}
\bibinfo{author}{\bibfnamefont{D.}~\bibnamefont{Goldhaber-Gordon}},
  \bibinfo{author}{\bibfnamefont{J.}~\bibnamefont{G{\"o}res}},
  \bibinfo{author}{\bibfnamefont{M.~A.} \bibnamefont{Kastner}},
  \bibinfo{author}{\bibfnamefont{H.}~\bibnamefont{Shtrikman}},
  \bibinfo{author}{\bibfnamefont{D.}~\bibnamefont{Mahalu}}, \bibnamefont{and}
  \bibinfo{author}{\bibfnamefont{U.}~\bibnamefont{Meirav}},
  \bibinfo{journal}{Phys. Rev. Lett.} \textbf{\bibinfo{volume}{81}},
  \bibinfo{pages}{5225} (\bibinfo{year}{1998}).

\bibitem[{\citenamefont{Kouwenhoven et~al.}(1997)\citenamefont{Kouwenhoven,
  Marcus, McEuen, Tarucha, Westervelt, and Wingreen}}]{kouwenhoven1997}
\bibinfo{author}{\bibfnamefont{L.~P.} \bibnamefont{Kouwenhoven}},
  \bibinfo{author}{\bibfnamefont{C.~M.} \bibnamefont{Marcus}},
  \bibinfo{author}{\bibfnamefont{P.~L.} \bibnamefont{McEuen}},
  \bibinfo{author}{\bibfnamefont{S.}~\bibnamefont{Tarucha}},
  \bibinfo{author}{\bibfnamefont{R.~M.} \bibnamefont{Westervelt}},
  \bibnamefont{and} \bibinfo{author}{\bibfnamefont{N.~S.}
  \bibnamefont{Wingreen}}, \emph{\bibinfo{title}{Proceedings of the NATO
  Advanced Study Institute on Mesoscopic Electron Transport, edited by L.L.
  Sohn, L.P. Kouwenhoven, and G. Sch{\"o}n}}, E345
  (\bibinfo{publisher}{Kluwer}, \bibinfo{year}{1997}).

\bibitem[{\citenamefont{Hanson et~al.}(2007)\citenamefont{Hanson, Kouwenhoven,
  Petta, Tarucha, and Vandersypen}}]{hanson2007}
\bibinfo{author}{\bibfnamefont{R.}~\bibnamefont{Hanson}},
  \bibinfo{author}{\bibfnamefont{L.~P.} \bibnamefont{Kouwenhoven}},
  \bibinfo{author}{\bibfnamefont{J.~R.} \bibnamefont{Petta}},
  \bibinfo{author}{\bibfnamefont{S.}~\bibnamefont{Tarucha}}, \bibnamefont{and}
  \bibinfo{author}{\bibfnamefont{L.~M.~K.} \bibnamefont{Vandersypen}},
  \bibinfo{journal}{Rev. Mod. Phys.} \textbf{\bibinfo{volume}{79}},
  \bibinfo{pages}{1217} (\bibinfo{year}{2007}).

\bibitem[{\citenamefont{Gerland et~al.}(2000)\citenamefont{Gerland, von Delft,
  Costi, and Oreg}}]{gerland2000}
\bibinfo{author}{\bibfnamefont{U.}~\bibnamefont{Gerland}},
  \bibinfo{author}{\bibfnamefont{J.}~\bibnamefont{von Delft}},
  \bibinfo{author}{\bibfnamefont{T.~A.} \bibnamefont{Costi}}, \bibnamefont{and}
  \bibinfo{author}{\bibfnamefont{Y.}~\bibnamefont{Oreg}},
  \bibinfo{journal}{Phys. Rev. Lett.} \textbf{\bibinfo{volume}{84}},
  \bibinfo{pages}{3710} (\bibinfo{year}{2000}).

\bibitem[{\citenamefont{Nozi\`eres}(1974)}]{nozieres1974}
\bibinfo{author}{\bibfnamefont{P.}~\bibnamefont{Nozi\`eres}},
  \bibinfo{journal}{J. Low Temp. Phys.} \textbf{\bibinfo{volume}{17}},
  \bibinfo{pages}{31} (\bibinfo{year}{1974}).

\bibitem[{\citenamefont{Pustilnik and Glazman}(2004)}]{pustilnik2004}
\bibinfo{author}{\bibfnamefont{M.}~\bibnamefont{Pustilnik}} \bibnamefont{and}
  \bibinfo{author}{\bibfnamefont{L.}~\bibnamefont{Glazman}},
  \bibinfo{journal}{Journal of Physics-Condensed Matter}
  \textbf{\bibinfo{volume}{16}}, \bibinfo{pages}{R513} (\bibinfo{year}{2004}).

\bibitem[{\citenamefont{Silvestrov and Imry}(2003)}]{silvestrov2003}
\bibinfo{author}{\bibfnamefont{P.~G.} \bibnamefont{Silvestrov}}
  \bibnamefont{and} \bibinfo{author}{\bibfnamefont{Y.}~\bibnamefont{Imry}},
  \bibinfo{journal}{Phys. Rev. Lett.} \textbf{\bibinfo{volume}{90}},
  \bibinfo{pages}{106602} (\bibinfo{year}{2003}).

\bibitem[{\citenamefont{Avinun-Kalish et~al.}(2005)\citenamefont{Avinun-Kalish,
  Heiblum, Zarchin, Mahalu, and Umansky}}]{mak2005}
\bibinfo{author}{\bibfnamefont{M.}~\bibnamefont{Avinun-Kalish}},
  \bibinfo{author}{\bibfnamefont{M.}~\bibnamefont{Heiblum}},
  \bibinfo{author}{\bibfnamefont{O.}~\bibnamefont{Zarchin}},
  \bibinfo{author}{\bibfnamefont{D.}~\bibnamefont{Mahalu}}, \bibnamefont{and}
  \bibinfo{author}{\bibfnamefont{V.}~\bibnamefont{Umansky}},
  \bibinfo{journal}{Nature} \textbf{\bibinfo{volume}{436}},
  \bibinfo{pages}{529} (\bibinfo{year}{2005}).

\bibitem[{\citenamefont{Hackenbroich and
  Weidenm{\"u}ller}(1996)}]{hackenbroich1996}
\bibinfo{author}{\bibfnamefont{G.}~\bibnamefont{Hackenbroich}}
  \bibnamefont{and} \bibinfo{author}{\bibfnamefont{H.~A.}
  \bibnamefont{Weidenm{\"u}ller}}, \bibinfo{journal}{Phys. Rev. Lett.}
  \textbf{\bibinfo{volume}{76}}, \bibinfo{pages}{110} (\bibinfo{year}{1996}).

\bibitem[{\citenamefont{Ji et~al.}(2000)\citenamefont{Ji, Heiblum, Sprinzak,
  Mahalu, and Shtrikman}}]{ji2000}
\bibinfo{author}{\bibfnamefont{Y.}~\bibnamefont{Ji}},
  \bibinfo{author}{\bibfnamefont{M.}~\bibnamefont{Heiblum}},
  \bibinfo{author}{\bibfnamefont{D.}~\bibnamefont{Sprinzak}},
  \bibinfo{author}{\bibfnamefont{D.}~\bibnamefont{Mahalu}}, \bibnamefont{and}
  \bibinfo{author}{\bibfnamefont{H.}~\bibnamefont{Shtrikman}},
  \bibinfo{journal}{Science} \textbf{\bibinfo{volume}{290}},
  \bibinfo{pages}{779} (\bibinfo{year}{2000}).

\bibitem[{\citenamefont{Ji et~al.}(2002)\citenamefont{Ji, Heiblum, and
  Shtrikman}}]{ji2002}
\bibinfo{author}{\bibfnamefont{Y.}~\bibnamefont{Ji}},
  \bibinfo{author}{\bibfnamefont{M.}~\bibnamefont{Heiblum}}, \bibnamefont{and}
  \bibinfo{author}{\bibfnamefont{H.}~\bibnamefont{Shtrikman}},
  \bibinfo{journal}{Phys. Rev. Lett.} \textbf{\bibinfo{volume}{88}},
  \bibinfo{pages}{76601} (\bibinfo{year}{2002}).

\bibitem[{\citenamefont{Sato et~al.}(2005)\citenamefont{Sato, Aikawa,
  Kobayashi, Katsumoto, and Iye}}]{sato2005}
\bibinfo{author}{\bibfnamefont{M.}~\bibnamefont{Sato}},
  \bibinfo{author}{\bibfnamefont{H.}~\bibnamefont{Aikawa}},
  \bibinfo{author}{\bibfnamefont{K.}~\bibnamefont{Kobayashi}},
  \bibinfo{author}{\bibfnamefont{S.}~\bibnamefont{Katsumoto}},
  \bibnamefont{and} \bibinfo{author}{\bibfnamefont{Y.}~\bibnamefont{Iye}},
  \bibinfo{journal}{Phys. Rev. Lett.} \textbf{\bibinfo{volume}{95}},
  \bibinfo{pages}{66801} (\bibinfo{year}{2005}).

\bibitem[{\citenamefont{Hackenbroich}(2001)}]{hackenbroich2001}
\bibinfo{author}{\bibfnamefont{G.}~\bibnamefont{Hackenbroich}},
  \bibinfo{journal}{Phys. Rep.} \textbf{\bibinfo{volume}{343}},
  \bibinfo{pages}{463} (\bibinfo{year}{2001}).

\bibitem[{\citenamefont{Ciorga et~al.}(2000)\citenamefont{Ciorga, Sachrajda,
  Hawrylak, Gould, Zawadzki, Jullian, Feng, and Wasilewski}}]{ciorga2000}
\bibinfo{author}{\bibfnamefont{M.}~\bibnamefont{Ciorga}},
  \bibinfo{author}{\bibfnamefont{A.~S.} \bibnamefont{Sachrajda}},
  \bibinfo{author}{\bibfnamefont{P.}~\bibnamefont{Hawrylak}},
  \bibinfo{author}{\bibfnamefont{C.}~\bibnamefont{Gould}},
  \bibinfo{author}{\bibfnamefont{P.}~\bibnamefont{Zawadzki}},
  \bibinfo{author}{\bibfnamefont{S.}~\bibnamefont{Jullian}},
  \bibinfo{author}{\bibfnamefont{Y.}~\bibnamefont{Feng}}, \bibnamefont{and}
  \bibinfo{author}{\bibfnamefont{Z.}~\bibnamefont{Wasilewski}},
  \bibinfo{journal}{Phys. Rev. B} \textbf{\bibinfo{volume}{61}},
  \bibinfo{pages}{R16315} (\bibinfo{year}{2000}).

\bibitem[{\citenamefont{Vidan et~al.}(2006)\citenamefont{Vidan, Stopa,
  Westervelt, Hanson, and Gossard}}]{vidan2006}
\bibinfo{author}{\bibfnamefont{A.}~\bibnamefont{Vidan}},
  \bibinfo{author}{\bibfnamefont{M.}~\bibnamefont{Stopa}},
  \bibinfo{author}{\bibfnamefont{R.~M.} \bibnamefont{Westervelt}},
  \bibinfo{author}{\bibfnamefont{M.}~\bibnamefont{Hanson}}, \bibnamefont{and}
  \bibinfo{author}{\bibfnamefont{A.~C.} \bibnamefont{Gossard}},
  \bibinfo{journal}{Phys. Rev. Lett.} \textbf{\bibinfo{volume}{96}},
  \bibinfo{pages}{156802} (\bibinfo{year}{2006}).

\bibitem[{\citenamefont{Yacoby et~al.}(1995)\citenamefont{Yacoby, Heiblum,
  Mahalu, and Shtrikman}}]{yacoby1995}
\bibinfo{author}{\bibfnamefont{A.}~\bibnamefont{Yacoby}},
  \bibinfo{author}{\bibfnamefont{M.}~\bibnamefont{Heiblum}},
  \bibinfo{author}{\bibfnamefont{D.}~\bibnamefont{Mahalu}}, \bibnamefont{and}
  \bibinfo{author}{\bibfnamefont{H.}~\bibnamefont{Shtrikman}},
  \bibinfo{journal}{Phys. Rev. Lett.} \textbf{\bibinfo{volume}{74}},
  \bibinfo{pages}{4047} (\bibinfo{year}{1995}).

\bibitem[{\citenamefont{Schuster et~al.}(1997)\citenamefont{Schuster, Buks,
  Heiblum, Mahalu, Umansky, and Shtrikman}}]{schuster1997}
\bibinfo{author}{\bibfnamefont{R.}~\bibnamefont{Schuster}},
  \bibinfo{author}{\bibfnamefont{E.}~\bibnamefont{Buks}},
  \bibinfo{author}{\bibfnamefont{M.}~\bibnamefont{Heiblum}},
  \bibinfo{author}{\bibfnamefont{D.}~\bibnamefont{Mahalu}},
  \bibinfo{author}{\bibfnamefont{V.}~\bibnamefont{Umansky}}, \bibnamefont{and}
  \bibinfo{author}{\bibfnamefont{H.}~\bibnamefont{Shtrikman}},
  \bibinfo{journal}{Nature} \textbf{\bibinfo{volume}{385}},
  \bibinfo{pages}{417} (\bibinfo{year}{1997}).

\bibitem[{\citenamefont{Sprinzak et~al.}(2002)\citenamefont{Sprinzak, Ji,
  Heiblum, Mahalu, and Shtrikman}}]{sprinzak2002}
\bibinfo{author}{\bibfnamefont{D.}~\bibnamefont{Sprinzak}},
  \bibinfo{author}{\bibfnamefont{Y.}~\bibnamefont{Ji}},
  \bibinfo{author}{\bibfnamefont{M.}~\bibnamefont{Heiblum}},
  \bibinfo{author}{\bibfnamefont{D.}~\bibnamefont{Mahalu}}, \bibnamefont{and}
  \bibinfo{author}{\bibfnamefont{H.}~\bibnamefont{Shtrikman}},
  \bibinfo{journal}{Phys. Rev. Lett.} \textbf{\bibinfo{volume}{88}},
  \bibinfo{pages}{176805} (\bibinfo{year}{2002}).

\bibitem[{\citenamefont{K{\"o}nig and Gefen}(2002)}]{konig2002}
\bibinfo{author}{\bibfnamefont{J.}~\bibnamefont{K{\"o}nig}} \bibnamefont{and}
  \bibinfo{author}{\bibfnamefont{Y.}~\bibnamefont{Gefen}},
  \bibinfo{journal}{Phys. Rev. B} \textbf{\bibinfo{volume}{65}},
  \bibinfo{pages}{45316} (\bibinfo{year}{2002}).

\end{thebibliography}

\end{document}